\documentclass[aps,prx,twocolumn,superscriptaddress,longbibliography]{revtex4-1}

\usepackage{amsmath}
\usepackage{amssymb}
\usepackage{amstext}
\usepackage{amsopn}
\usepackage{amsfonts}
\usepackage{amsxtra}
\usepackage[english]{babel}
\usepackage{graphicx}
\usepackage{bm}
\usepackage{multirow}
\usepackage{dcolumn}
\usepackage{color}
\usepackage{hyperref}
\usepackage{todonotes}
\usepackage{verbatim}
\begin{document}

\title{Formation of Incommensurate Charge Density Waves in Cuprates} 

\author{H. Miao}\email[]{hmiao@bnl.gov}
\affiliation{Condensed Matter Physics and Materials Science Department, Brookhaven National Laboratory, Upton, New York 11973, USA}

\author{R. Fumagalli}
\affiliation{Dipartimento di Fisica, Politecnico di Milano, Piazza Leonardo da Vinci 32, 20133 Milano, Italy}
\author{M. Rossi}
\affiliation{Dipartimento di Fisica, Politecnico di Milano, Piazza Leonardo da Vinci 32, 20133 Milano, Italy}
\affiliation{Stanford Institute for Materials and Energy Sciences, SLAC National Accelerator Laboratory and Stanford University, 2575 Sand Hill Road, Menlo Park, California 94025, USA}

\author{J. Lorenzana}
\affiliation{ISC-CNR, Dipartimento di Fisica, Universit\`a di Roma ``La Sapienza'', Piazzale Aldo Moro, 00185 Roma, Italy}
\author{G. Seibold}
\affiliation{Institut f\"{u}r Physik, BTU Cottbus, P.O.\ Box 101344, 03013 Cottbus, Germany}

\author{F. Yakhou-Harris}
\author{K. Kummer}
\author{N. B. Brookes}
\affiliation{European Synchrotron Radiation Facility (ESRF), BP 220, F-38043 Grenoble Cedex, France}

\author{G. D. Gu}
\affiliation{Condensed Matter Physics and Materials Science Department, Brookhaven National Laboratory, Upton, New York 11973, USA}
\author{L. Braicovich}
\affiliation{Dipartimento di Fisica, Politecnico di Milano, Piazza Leonardo da Vinci 32, 20133 Milano, Italy}
\affiliation{European Synchrotron Radiation Facility (ESRF), BP 220, F-38043 Grenoble Cedex, France}
\author{G. Ghiringhelli}
\affiliation{Dipartimento di Fisica, Politecnico di Milano, Piazza Leonardo da Vinci 32, 20133 Milano, Italy}
\affiliation{CNR/SPIN, Politecnico di Milano, Piazza Leonardo da Vinci 32, 20133 Milano, Italy}

\author{M. P. M. Dean}\email[]{mdean@bnl.gov}
\affiliation{Condensed Matter Physics and Materials Science Department, Brookhaven National Laboratory, Upton, New York 11973, USA}

\date{\today}



\date{\today}

\begin{abstract}
Although charge density waves (CDWs) are omnipresent in cuprate high-temperature superconductors, they occur at significantly different wavevectors, confounding efforts to understand their formation mechanism. Here, we use resonant inelastic x-ray scattering to investigate the doping- and temperature-dependent CDW evolution in La$_{2-x}$Ba$_{x}$CuO$_{4}$ ($x=0.115-0.155$). We discovered that the CDW develops in two stages with decreasing temperature. A precursor CDW with quasi-commensurate wavevector emerges first at high-temperature. This doping-independent precursor CDW correlation originates from the CDW phase mode coupled with a phonon and ``seeds" the low-temperature CDW with strongly doping dependent wavevector. Our observation reveals the precursor CDW and its phase mode as the building blocks of the highly intertwined electronic ground state in the cuprates.   
\end{abstract}

\maketitle

\begin{figure*}[tb]
\includegraphics[width=10.5 cm]{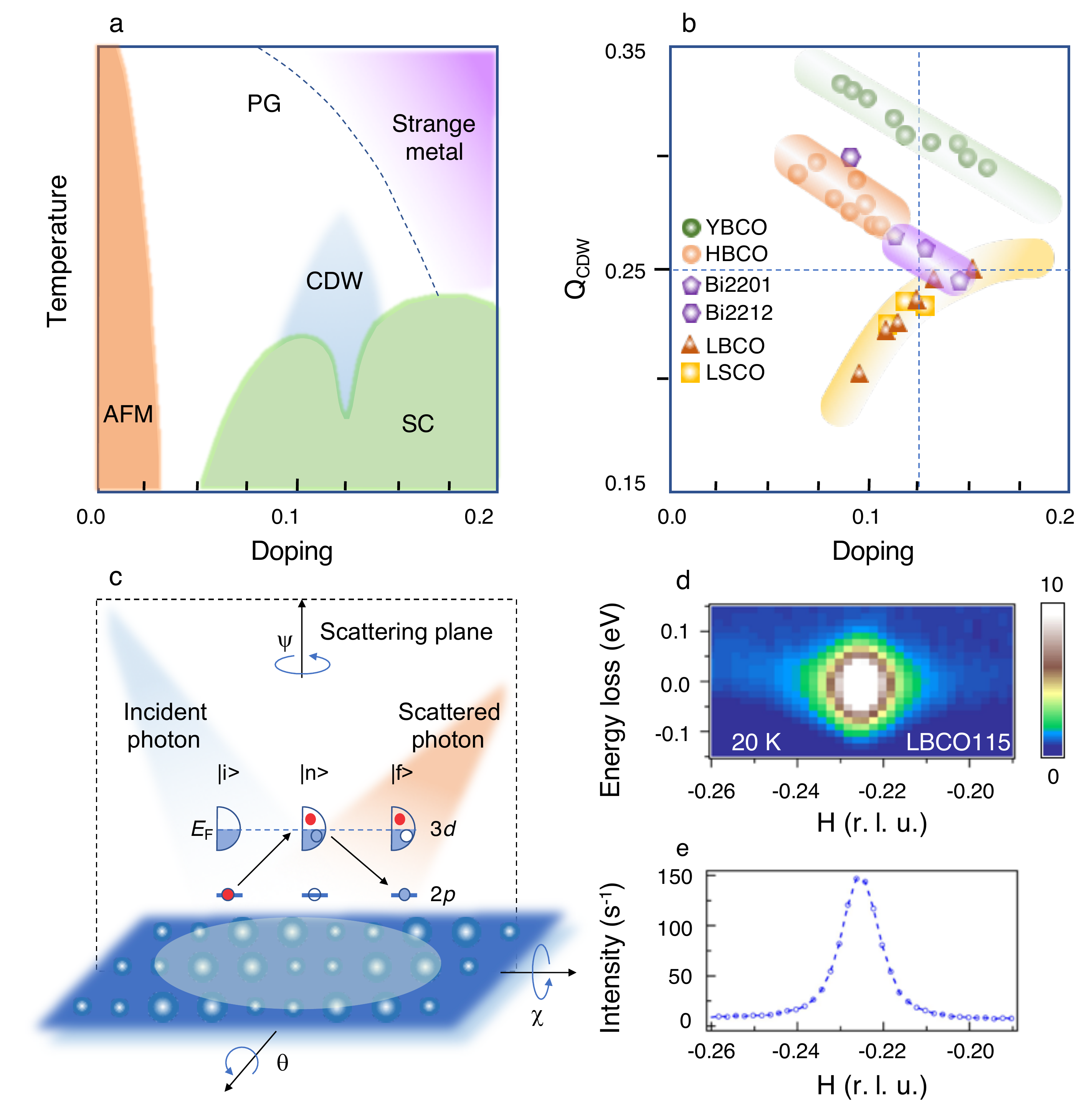}
\caption{\textbf{CDWs in the cuprates}. (a) Schematic phase diagram of the cuprates. A ubiquitous CDW dome is observed below the pseudogap temperature and coexists with superconductivity. (b) The doping-dependent CDW wavevector, $\mathbf{Q}_{\text{CDW}}$, in various cuprate families at low temperature \cite{Hucker2011,Blanco-Canosa2014,Tabis2017,Comin2014,Chaix2017, Miao2017, Miao2018}. (c) Schematically shows the \textit{L}-edge RIXS process and experimental setup. $|i\rangle$, $|n\rangle$ and $|f\rangle$ represent initial, intermediate and final states, respectively. Solid and empty circles represent occupied and unoccupied states, respectively. (d) Typical RIXS intensity map of La$_{2-x}$Ba$_{x}$CuO$_{4}$ ($x=0.115$) at 20~K. The momentum transfer is obtained by rotating the sample about the $\theta$ or $\chi$ axis. (e) Integrated RIXS intensity in a $\pm100$~meV energy window of (d) shows a strong CDW peak at $\mathbf{Q}_{\text{CDW}}=0.225$~r.l.u.}
\label{Fig1}
\end{figure*}

A remarkable phenomenon of the cuprates is the coexistence of multiple nearly-degenerate electronic orders or instabilities that intertwine at low temperature to form the novel electronic liquid which precipitates high-$T_{c}$ superconductivity \cite{Fradkin2015, Lee2014}. While unidirectional charge density waves (CDWs), also known as stripes, have been theoretically predicted for doped Mott insulators \cite{Zaanen1989, Poilblanc1989, Emery1990, Castellani1995} and experimentally discovered in La-based cuprates over two decades ago \cite{Tranquada1995}, a full CDW phase diagram for different cuprate systems, as shown schematically in Fig.~\ref{Fig1}(a), was only established very recently \cite{Hucker2011,Ghiringhelli2012,Blanco-Canosa2014,Tabis2017,Comin2014, Abbamonte2005, Croft2014, Thampy2014, Miao2017, Miao2018, Arpaia2018}. Consistent results are also found in state-of-the-art numerical calculations of realistic 2D $t-J$ and Hubbard models near $1/8$ doping, where stripe ordering or fluctuations are found to be one of the leading electronic instabilities of the ground state \cite{Corboz2014, Huang2017, Zheng2017}. While this progress indicates a universal CDW mechanism, consensus about the nature of this mechanism has not been reached due to the opposite evolution of the CDW wavevectors with doping in different cuprates families. Figure~\ref{Fig1}(b) summarizes the doping dependence of the CDW wavevector as determined by diffraction measurements \cite{Hucker2011,Blanco-Canosa2014,Tabis2017,Comin2014,Chaix2017, Miao2017, Miao2018}. In the La-based cuprates, such as La$_{2-x}$Ba$_{x}$CuO$_{4}$ (LBCO), CDW wavevectors increase with doping and saturate at doping levels beyond $x=0.125$. In the Bi-, Y- and Hg-based cuprates, however, CDW wavevectors monotonically decrease with doping. These observations have motivated different pictures for CDW formation mechanisms based on either real-space local interactions or weak coupling Fermi surface (FS) driven mechanisms \cite{Zaanen1989, Poilblanc1989, Tranquada1995, Comin2015, Shen2005}. However, as is evident from the incommensurate-commensurate crossover in transition metal chalcogenides \cite{Gruner2018}, low-temperature ordering wavevectors are not necessarily representative of the CDW formation mechanism, which is instead encoded in the inelastic spectrum above the transition temperature. As we show in Fig.~\ref{Fig1}c, resonant-inelastic x-ray scattering (RIXS) can probe electronic degrees of freedom via its resonant process  \cite{Ament2011, Dean2015}. Together with the improvement of energy resolution, RIXS can thus reveal CDW order and its fluctuations in great detail.


%

To understand the nature and formation mechanism of the CDW, we use RIXS to study the doping and temperature dependent CDW evolution in LBCOn (n=115, 125 and 155, corresponding to x=0.115, 0.125 and 0.155 in La$_{2-x}$Ba$_{x}$CuO$_{4}$, respectively).  By carefully tracing the doping and temperature dependent elastic and inelastic CDW signals in the RIXS spectra, we discovered that a doping-independent precursor CDW with quasi-commensurate wavevector is developed first at high temperature. This short-ranged CDW correlation originates from the phase mode of the CDW and ``seeds" the long-ranged CDW with strong doping-dependent incommensurate wavevectors at lower temperature. This two-stage CDW evolution uncovers the locally commensurate CDW together with its inelastic excitation as the building block of the charge correlations in the underdoped cuprates and suggests that the doping-dependent incommensurate CDW wavevectors are driven by the subtle balance of intertwined spin, charge and lattice correlations.  

%
\begin{figure*}[tb]
\includegraphics[width=17 cm]{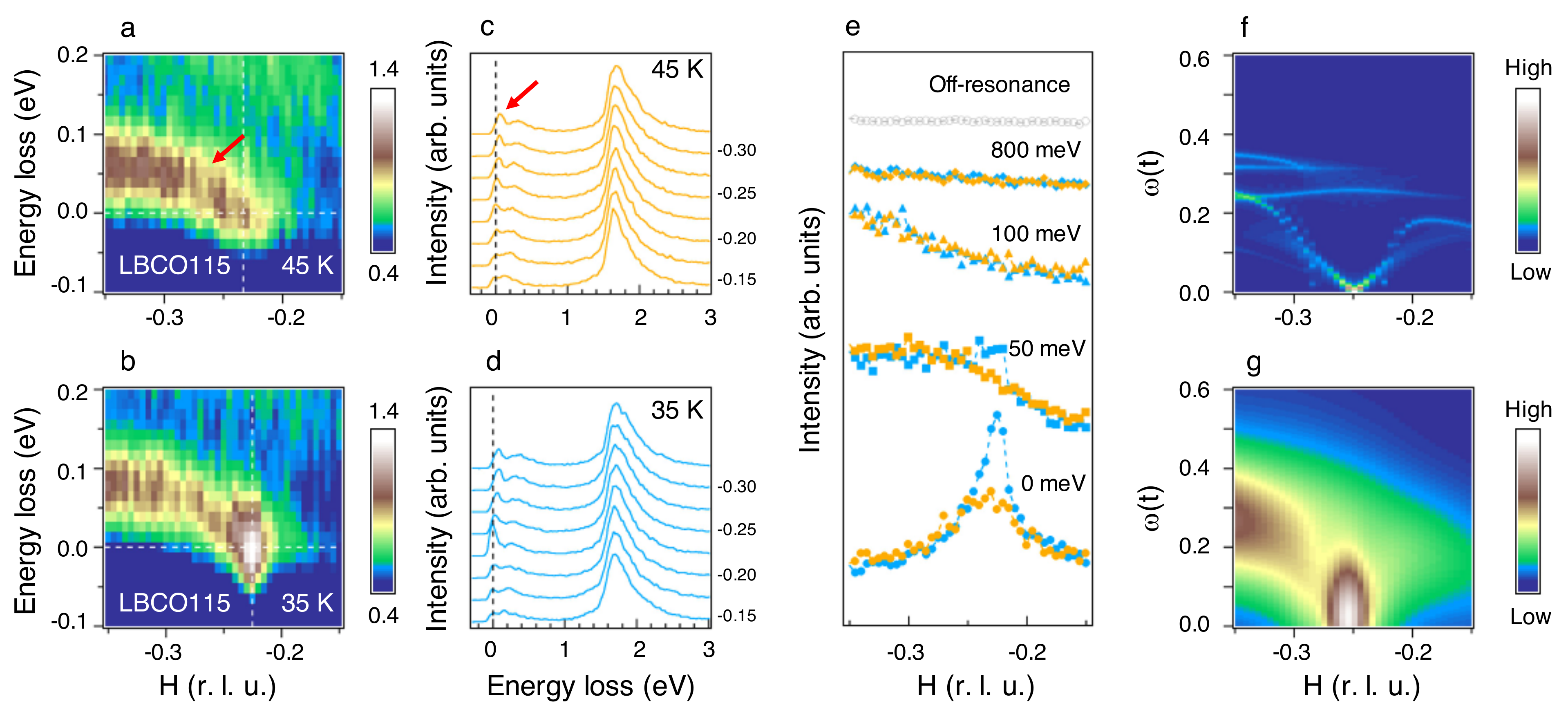}
\caption{\textbf{Two-stage CDW correlations}. (a) and (b) show RIXS intensity maps of LBCO115 in wide momentum transfer range at 45~K and 35~K, respectively. Note that the color scales are nearly 10 times smaller than in Fig.~\ref{Fig1}d. Representative constant $\mathbf{Q}$ cut of (a) and (b) are shown in (c) and (d), respectively. Constant $E_{\text{loss}}$ cut of (a) and (b) at $E_{\text{loss}}=$0, 50, 100 and 800~meV are shown in (e). The yellow and cyan colors represent data at 45 and 35~K, respectively. Gray squares represent an off-resonant constant $E_{\text{loss}}$ cut at 0~meV. Below 500~meV, due to the dispersive charge and paramagnon excitation, the intensity maximum is shifted to larger $\mathbf{Q}$. At 800~meV, the constant energy cut becomes flat and similar to non-resonant constant energy cut at zero energy loss. (f) Calculated charge susceptibility along the $(0,0)\rightarrow(-$H$,0)$ direction. The $d=4$ stripe phase mode couples with phonon. (g) Same plot as (f) but convoluted with experimental resolution. The energy axes in (f) and (g) are shown in units of $t \approx 0.3$~eV.}
\label{Fig2}
\end{figure*}

We start by revealing the two-stage CDW evolution in LBCO115. Figures~\ref{Fig1}(d) and (e) show a typical RIXS intensity plot and the integrated RIXS intensity ($\pm100$~meV with respect to zero energy loss) at 20~K. The strong intensity centered at zero-energy and $|\mathbf{Q}|=0.225$ in reciprocal lattice units (r.l.u.) corresponds to static CDW order in LBCO115. RIXS intensity plots of LBCO115 below 200~meV at 45 and 35~K are shown in Figs.~\ref{Fig2}(a) and (b), respectively. Representative constant momentum spectra in a wider energy range are shown in Figs.~\ref{Fig2}(c) and (d), where the well-established dispersionless $dd$ excitations ($\sim1.7$~eV) and dispersive paramagnon ($200\sim 350$~meV) are observed \cite{Miao2017,Dean2012,Ghiringhelli2012, LeTacon2011}. At 45~K, the RIXS spectra below 100~meV are dominated by dispersive charge excitations (identified by red arrows in Fig.~\ref{Fig2}a and c) whose intensity quickly fades away below $|\mathbf{Q}|\sim 0.2$~r.l.u. This new feature was not observed in previous RIXS studies of this system, due to poorer energy resolution \cite{DeanLBCO2013, Miao2017}.  A zero-energy (0~meV) cut of the intensity plot shows a broad quasielastic peak along the $H$ direction (yellow curve at the bottom of Fig.~\ref{Fig2}(e)) hereafter referred to as the precursor-CDW peak (pCDW). At higher energy, the peak position of the constant energy cut shifts to higher $\mathbf{Q}$ which may affect the $\mathbf{Q}_{\text{pCDW}}$ in energy integrated diffraction study. This broad peak intensity is completely suppressed when changing the incident photon energy 1.5~eV below the Cu $L_{3}$ edge (gray curve in Fig.~\ref{Fig2}(e)), thus proving that the signal is dominated by the resonant process. The large inelastic contribution and broad peak width of the pCDW suggest dynamic charge fluctuations as discussed extensively in a different cuprate family recently \cite{Arpaia2018}. Intriguingly, Fig.~\ref{Fig2}(e) shows that $H$ cuts at 50 and 100~meV show stronger spectral weight at larger values of $|H|$ indicating that dynamic charge correlations may tend to exist at higher $|H|$. It is these higher-energy dynamic correlations that drive the motion of the total energy-integrated CDW peak, and the associated phonon softening, to $H=0.272$~r.l.u.\ at higher temperatures of 90~K, although the worse energy resolution of the previous RIXS measurements was insufficient to separate out this effect \cite{Miao2017, Miao2018}. As we cool down to 35~K an elastic peak emerges on top of the broad dispersive feature and eventually evolves to the intense CDW peak shown in Fig.~\ref{Fig1}(d). To distinguish these two CDW peaks, we refer to the low temperature peak as the low temperature-CDW (lCDW).

To understand the origin of the inelastic excitation and its connection with the CDW, we calculated the dynamic charge susceptibility, $\chi^{ee}_{\mathbf{q}}(\omega)$, of a phenomenological model that reproduces our observations. This assumes the presence of metallic stripes within a correlated Hubbard model at low temperatures, since phonons are known to have large contributions in the energy range of interest \cite{Chaix2017,Devereaux2016, Reznik2006}, we also include a phonon mode of energy $\Omega_{\mathbf{q}}$, which couples to the electrons with interaction vertex, $g_{\mathbf{q}}$. Figure~\ref{Fig2}(f) and (g) show the calculated spectra and the experimental resolution convoluted spectra, respectively. We choose parameters so that the phase mode of the CDW yields an acoustic mode dispersing out from $\mathbf{Q}_{\text{CDW}}=0.25$~r.l.u.\ and interacts strongly with the phonon mode at low $Q$. This regime of soft phasons was invoked before to explain the optical conductivity \cite{Lorenzana2003a}. Here, the phonon-phason coupling yields the large momentum dependence of the inelastic intensity observed in Fig.~\ref{Fig2}(a) and (b). 

Figure~\ref{Fig2}(g) shows that the model reproduces quite well the features observed at low temperatures even though disorder is neglected. The sensitivity of a CDW to disorder is dictated by its stiffness to local phase changes i.e.\ the energy cost to distorting the CDW phase locally, so that it can pin to a point defect \cite{Gruner2018}. A stiff CDW will tend to preserve its local phase and will therefore be inefficiently pinned by disorder; whereas a flexible CDW will distort such that it efficiently pinned. The high-temperature signal is consistent with a flexible pCDW that is strongly pinned by disorder while the low-temperature features can be assigned to a small fraction of the CDW which becomes stiff and is therefore inefficiently pinned. This can be seen by noting that the total $q$-integrated scattering from the pCDW is 7 times larger than that from the lCDW \cite{Miao2017}. Such a phenomenology explains the concomitant presence of long-range charge order and a well-defined phason mode due to poor pinning. It is worth emphasizing that in the pCDW state the phason mode is still clearly present but yields a broad structure at low energy. 

%
\begin{figure*}[tb]
\includegraphics[width=13 cm]{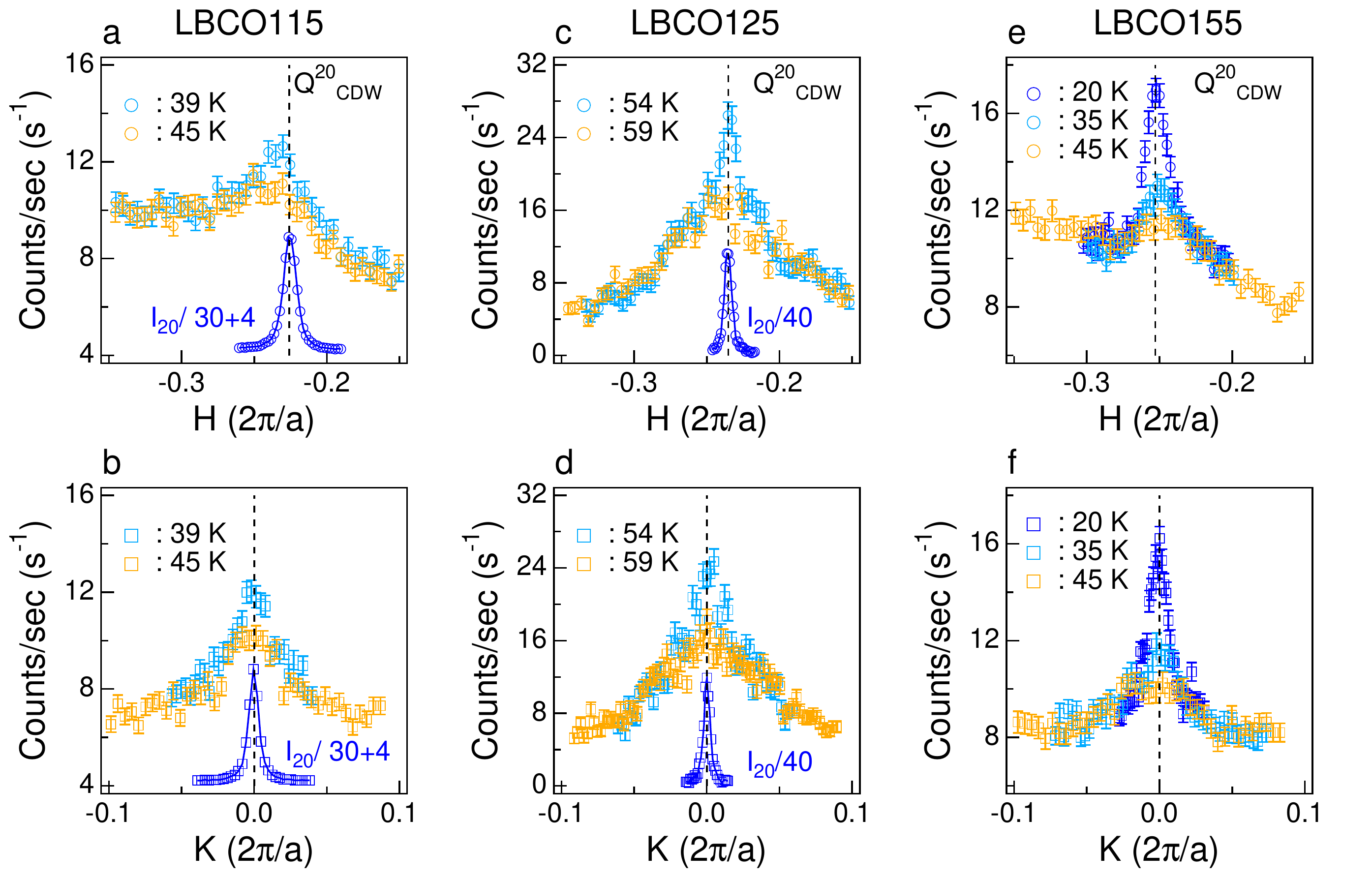}
\caption{\textbf{The doping dependent evolution of the two-stage CDW}. (a) and (b) show integrated RIXS intensity ($\pm$100~meV with respect to the elastic line) of LBCO115 along the $H$ and $K$ directions, respectively. The same plots of LBCO125 and LBCO155 are shown in (c),(d) and (e),(f) respectively. Blue symbols represent data at 20~K. Due to the large lCDW signal in LBCO115 and LBCO125, the 20~K intensity, $I_{20}$, is normalized and offset to $I_{20}/3+4$ and $I_{20}/4$ for LBCO115 and LBCO125, respectively. The cyan and yellow symbols represent the data just below and just above the critical temperature, where the lCDW starts to emerge. Dashed lines represent $\mathbf{Q}_{\text{CDW}}$ at 20~K.}
\label{Fig3}
\end{figure*}

We now explore the doping-dependent evolution of the two-stage CDW, as enabled by higher RIXS throughput \cite{Brookes2018beamline}. In Fig.~\ref{Fig3}, we show the quasielastic RIXS intensity of LBCO115, LBCO125 and LBCO155 along the $H$ and $K$ directions. This is obtained by integrating $\pm100$~meV respect to the elastic line in order to achieve higher sensitivity than cuts at 0~meV.  At 20~K (blue symbols), the lCDW peaks are strongly doping-dependent. The peak intensity is largest in LBCO125 and significantly weaker in LBCO155, consistent with the lCDW dome centered at $1/8$ doping \cite{Hucker2011, Blanco-Canosa2013, Tabis2017}. A similar trend is shown in the correlation length, $\xi$, defined as the inverse peak half-width-at-half-maximum (1/HWHM), that is largest in LBCO125 and shortest in LBCO155. The peak position, $\mathbf{Q}_{\text{lCDW}}$, increases with doping and saturates for $x>0.125$. Here $x$ is the hole doping. As we warm up, the intensity of the lCDW decreases and disappears at $41\pm$2~K, $55\pm1$~K, $43\pm3$~K in LBCO115, LBCO125 and LBCO155, respectively. Near these critical temperatures, the two-stage CDW structure is evident along both the $H$ and $K$ directions.  

%
\begin{figure*}[tb]
\includegraphics[width=11 cm]{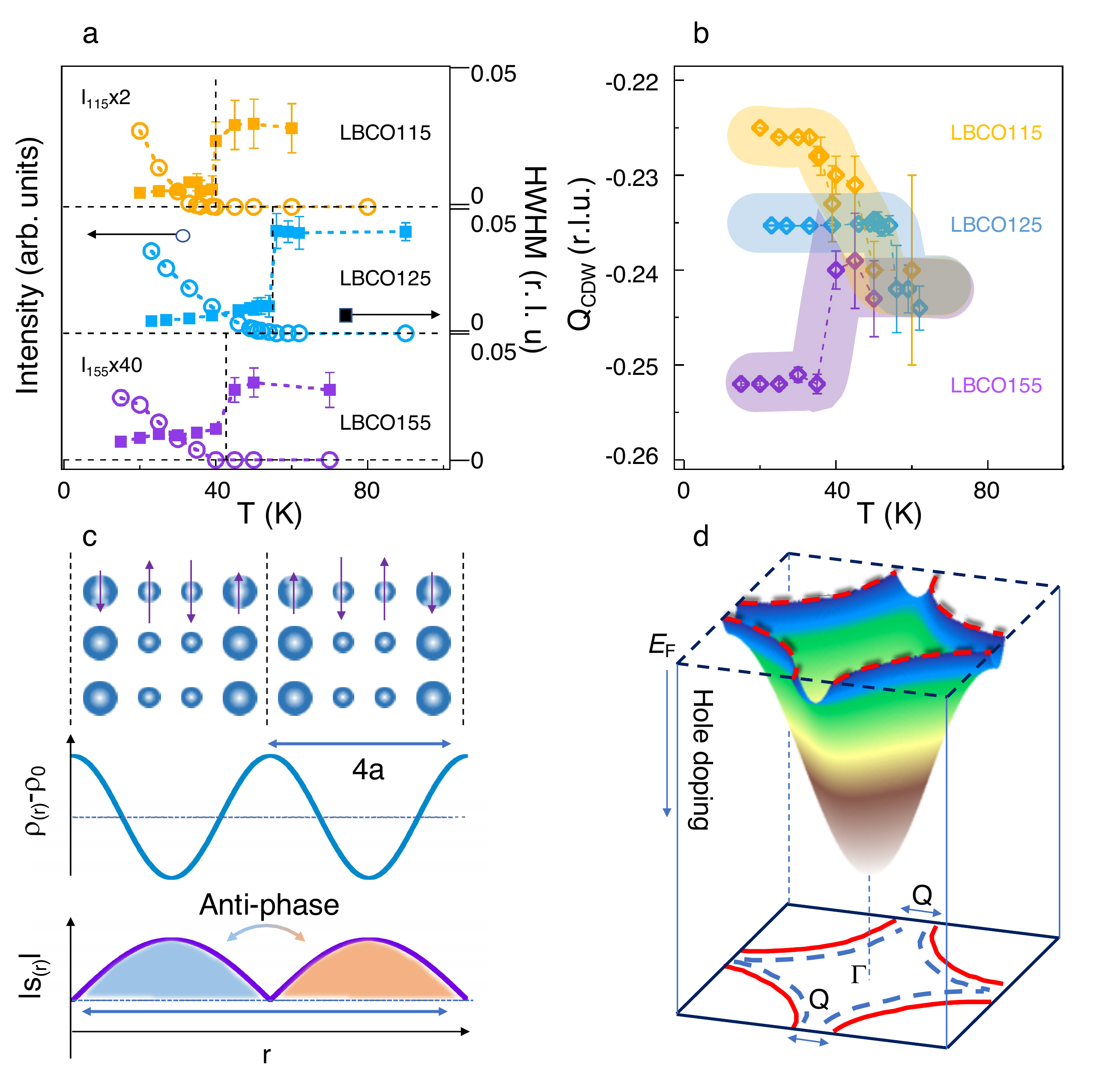}
\caption{\textbf{Universal pCDW}. The extracted peak intensity (left axis) and HWHM (righ axis) of the pCDW and lCDW are shown in (a). The vertical dashed lines at 38, 55 and 40~K represent the lCDW critical temperature of LBCO115, LBCO125 and LBCO155, respectively. Above these temperatures, both the peak intensity and the HWHM remains unchanged within the error bars. (b) shows the extracted temperature dependent CDW wavevectors. Note that our measurement is performed at negative $\mathbf{Q}_{\text{pCDW}}$ to enhance charge excitations \cite{Miao2017, Ghiringhelli2012}. The shaded yellow, blue and purple curves are guide-to-the-eye for LBCO115, LBCO125 and LBCO155. In the pCDW phase, the wavevectors are doping independent. (c) schematically shows the ``stripe" picture from the mean field theory. Blue circles and purple arrows represent the local hole and spin density. Anti-phase SDW domains are separated by charge stripes and give rise to the $|\mathbf{Q}_{\text{CDW}}|=2\delta_{\text{SDW}}\sim2x$ relation. (d) schematically shows the FS driven mechanism. Since the FS shrinks at higher hole-doping, this model predicts smaller CDW wavevectors at higher doping.}
\label{Fig4}
\end{figure*}

To quantify the doping and temperature dependence of the two-stage CDWs, we summarize the fitted CDW peak intensity, HWHM and the wavevectors in Figs.~\ref{Fig4}(a) and (b). Most remarkably, as we show in Fig.~\ref{Fig4}(b), we discovered that while the wavevectors of the lCDW is strongly doping-dependent, the wavevectors of the pCDW are doping independent and broadly peaked at $|\mathbf{Q}_{\text{pCDW}}|=0.240$~r.l.u. The corresponding real space CDW period, $\lambda_{pCDW}=1/\mathbf{Q}_{\text{pCDW}}\sim16\AA$, is similar to the extracted correlation length of pCDW, $\xi_{\text{pCDW}}=$18(2), 13(2) and 21(3)\AA~for LBCO115, LBCO125 and LBCO155, respectively, and suggests the existence of locally commensurate correlations without extended phase coherence. This picture is also in agreement with our theoretical considerations pointing to a ``soft" pCDW that is pinned by disorder. As we go on to discuss, these observations have important implications for the CDW phenomena observed in underdoped cuprates. 

Following Fig.~\ref{Fig1}(b), the wavevector of the CDW appears to fall in two categories with distinct doping-dependent trends. Figure~\ref{Fig4}(c) illustrates the real space stripe CDW mechanism, where by locating holes at the anti-phase SDW domain boundaries, the kinetic energy of the strongly correlated electrons is reduced. In this picture, when both CDW and SDW are static, the CDW wavevector is expected to follow the SDW with a simple $|\mathbf{Q}_{\text{CDW}}|=2\delta_{\text{SDW}}$ relation that is observed in La-based cuprates at low temperature and reconfirmed in our RIXS study. When the SDW is dynamic with a spin gap and no magnetic Bragg peak, the CDW wavevector is expected to unlock from the spin correlations with nearly degenerate wavevectors \cite{Miao2017, Miao2018, Nie2017, Huang2017, Zheng2017}. In the FS-based mechanism, the CDW is determined by FS portions with large density-of-states (DOS), and the free energy is minimized by reducing the DOS near the FS. Since hole doping shifts the chemical potential down in Fig.~\ref{Fig4}(d), the CDW wavevectors are expected to decrease with doping as has been observed in Bi-, Y- and Hg-based cuprates. Our observations of the pCDW and its phase mode demonstrate that the intrinsic CDW correlations emerge first with doping independent quasi-commensurate periods. This strongly points towards models in which CDW order is driven by local real-space correlations. Similar short-ranged CDW correlations that persist even to room temperature were recently observed in YBa$_2$Cu$_3$O$_{6+\delta}$ (YBCO) \cite{Wu2015, Arpaia2018}, Bi$_{2}$Sr$_{2}$CaCu$_{2}$O$_{8+\delta}$ (Bi2212) \cite{Chaix2017}, La$_{2-x}$Sr$_{x}$CuO$_{4}$ (LSCO) \cite{Croft2014, Thampy2013} and electron doped Nd$_{2-x}$Ce$_{x}$CuO$_{4}$ \cite{Neto2018}, strongly indicating an ubiquitous pCDW phase in underdoped cuprates. Our results are also compatible with previous STM studies of various cuprate families without magnetic stripe order at low temperature, such as Bi2212 and Ca$_{2-x}$Na$_{x}$CuO$_{2}$Cl$_{2}$, where CDWs are found to be locally commensurate with large phase slips \cite{Hanaguri2004, Howald2003,Mesaros2016}. The CDW phase mode and the pCDW thus serve as the ``seed" of the lCDW that couples strongly to different types of correlations at low temperature and is dragged to distinct wavevectors. An important prerequisite of this picture is that CDW states with different period are close in free energy. This is indeed supported by early computations \cite{Lorenzana2002} and recent state-of-the-art numerical studies of realistic $t-J$ model and 2D Hubbard model near 1/8 doping \cite{Corboz2014, Huang2017, Zheng2017}, where multiple CDW periods are nearly degenerate in energy. It would be interesting and important for future studies to explore the pCDW and its phase mode in heavily underdoped and overdoped cuprates (e.g.~LSCO) and built its connections with the puzzling pseudogap and strange metal phase.  

Finally we discuss the temperature-dependent commensurability effect observed in Fig.~\ref{Fig4}(b). Similar effects have been observed in prototypical stripe ordered La$_{2-x}$Sr$_{x}$NiO$_{4}$ (LSNO, $x\sim1/3$). In these materials, the CDW wavevectors also follow a simple $|\mathbf{Q}_{\text{CDW}}|=\delta_{\text{SDW}}\sim x$ relation at low-temperature and move to $\mathbf{Q}_{\text{CDW}}=$1/3 at high temperature. An entropy-driven self-doping mechanism has been proposed to explain the commensurability effect in LSNO \cite{Ishizaka2004}. This model considers the entropy of doped holes as $S = k_{B}\text{ln}(N_{c}$) where $N_{c}$ is the number of configurations for a given concentration of holes. The number of configurations is computed as the number of ways to accommodate indistinguishable particles in boxes representing equivalent sites along the core of the domain wall that can accommodate holes. We expanded the entropy model to our case as described in Appendix \ref{entropy}. This requires the commonly applied assumption of ordered holes along the stripe so that half-filled stripes are insulating with zero entropy and satisfy $\mathbf{|Q|}=2 x$. At finite temperatures it is convenient either to increase or decrease the incommensurability with respect to the $2x$ value to gain entropy (see Fig.~\ref{Figure5}). For doping levels below $x=0.125$ the solution with larger incommensurability has lower free energy and the CDW is predicted to be at  
\begin{equation}
    \mathbf{|Q|}=\frac{\epsilon_0}{1-e^{-E_{\text{g}}/2k_{\text{B}}T}}  
\label{CDWentropy1}
\end{equation}
where $\epsilon_0\sim 2x$ is the low-temperature incommensurability and $E_{\text{g}}$ is the energy gap due to the secondary order along the stripe. For doping levels higher than $x=0.125$ the computation is more complicated because stripes overlap and inter-stripe interactions become important producing a saturation of the low temperature incommensurability \cite{Lorenzana2002}.
Qualitatively we expect that the solution in which the incommensurability 
decreases with temperature prevails yielding at low temperatures,
\begin{equation}
    \mathbf{|Q|}=\epsilon_0 (1-e^{-E_{\text{g}}/2k_{\text{B}}T}).  
\label{CDWentropy2}
\end{equation}
This simple computation predicts an activated increase (decrease) of the incommensurablity for $x\lessapprox 1/8$ ($x\gtrapprox 1/8$) as indeed found (see Fig.~\ref{Figure6}).

In summary, we report detailed measurements of the doping and temperature dependent CDW correlations in LBCO. We discovered that CDW order forms from a doping-independent pCDW with quasi-commensurate period and a soft phase mode. Our observation thus uncovers the basic foundation underpinning the emergence of CDW order in the cuprates.

\begin{acknowledgements}
H.M.\ and M.P.M.D.\ acknowledge V. Bisogni, J. Tranquada and I. Robinson for insightful discussions. This material is based upon work supported by the U.S. Department of Energy, Office of Basic Energy Sciences, Early Career Award Program under Award No. 1047478. Work at Brookhaven National Laboratory was supported by the U.S. Department of Energy, Office of Science, Office of Basic Energy Sciences, under Contract No. DE-SC00112704. RIXS measurements were performed at the ID32 beamline of the European Synchrotron Radiation Facility (ESRF). J.L.\ acknowledges financial support from Italian MAECI through projects SUPERTOP-PGR04879 and AR17MO7, from MIUR though project PRIN 2017Z8TS5B and from Regione Lazio (L.R.\ 13/08) under project SIMAP.
\end{acknowledgements}

\appendix 

\section{Methods}
La$_{2-x}$Ba$_{x}$CuO$_{4}$ ($x$=0.115-0.155) single crystals were grown using the floating zone method and cleaved in-situ to reveal a face with a $[001]$ surface normal. The wavevectors used here are described using the high temperature tetragonal ($I4/mmm$) space group. The orientation matrix is determined by (002), (101) and (-101) fundamental peaks at 1700~eV.

RIXS measurements were performed at the ID32 beamline of the European Synchrotron Radiation Facility (ESRF). The resonant condition was achieved by tuning the incident x-ray energy to the maximum of the Cu $L_{3}$ absorption peak around 931.5~eV. The scattering geometry is shown in Fig.~\ref{Fig1}c. $\sigma$ and $\pi$ x-ray polarizations are defined as perpendicular and parallel to the scattering plane, respectively. $H$ and $K$ scans are achieved by rotating the sample around the $\theta$ and $\chi$ axes, without changing $2\theta$, thus changing the in-plane component of the momentum transfer $\mathbf{Q}$ = $\mathbf{k}_{f}$-$\mathbf{k}_{i}$. By doing this, we are assuming that the scattering is independent of $L$, which is reasonable as the inter-layer coupling in the cuprates is known to be weak \cite{Wilkins2011, Hucker2011, DeanLSCO2013}. All intensities are normalized to beam current and counting time. In this study, we used $\sigma$-polarized incident x-rays and negative $H$ values to enhance charge excitations \cite{Ghiringhelli2012, Miao2017}. In principle one can use the polarization analyzer at ID32 to ensure that  the excitation is a pure charge mode ($\Delta S=0$) \cite{Neto2018}. However, the efficiency of this setup is an order of magnitude lower than the standard setup which makes its use very time consuming for the present problem.  As a consequence we cannot completely exclude a spin flip component of the dispersing mode, although we consider it very unlikely because of the association of the mode with the charge
quasi-elastic scattering. The spectrometer scattering angle (2$\theta$) was fixed at $118^{\circ}$ such that $L\approx 1.5$ and the total instrumental energy resolution (full-width at half maximum) was set to 70~meV to increase the counting rate.  The quasi-elastic intensity was obtained by integrating the RIXS spectrum in an energy window of $\pm 100$~meV around 0~meV. 

\section{Charge excitations of stripes coupled to phonons} 

Our calculations are based on the single-band Hubbard model
\begin{equation}\label{eq:hub}
  H=\sum_{i,j,\sigma}t_{ij}c_{i,\sigma}^{\dagger}c_{j,\sigma}
  +U\sum_i n_{i,\uparrow}n_{i,\downarrow}
\end{equation}
where we include nearest ($\sim t$) and next-nearest neighbor ($\sim t'$) hopping. Stripe solutions are evaluated within Hartree-Fock (HF) and we calculate binding energies with respect to the homogeneous antiferromagnet (AF) for a configuration where the domain wall of the AF order parameter is bond-centered. Within the HF approximation site-centered stripes involve paramagnetic sites with charge density $n$ and the associated energy cost $\sim U n^2/4$ makes them energetically unfavorable with respect to bond-centered configurations for large $U$. This is not anymore the case if correlations beyond the HF approximation are taken into account\cite{Seibold2004b}. Here, for simplicity, we keep the HF approximation but choose parameters $U/t=4$ and $t'/t=-0.25$ which reproduce the ``Yamada-plot" \cite{Yamada1998}, i.e.\ the low-temperature relation between spin incommensurability $\delta_{\text{SDW}}$ and doping,  $x=\delta_{\text{SDW}}$ (see  Fig.~\ref{Fig4}(c)).

Excitations on top of the mean-field stripes are computed with the random phase approximation (RPA). The striped ground state couples charge fluctuations $\delta\rho_{\bf q}$ which differ by multiples $\alpha$ of the stripe modulation wave-vector ${\bf Q}_{s}$ with $\alpha$ an integer. Moreover charge fluctuations are coupled with fluctuations of the magnetization $\delta m_{{\bf q}+\alpha{\bf Q}_s}$ so that for each $\alpha$, $\beta$ the susceptibility  is a $2\times 2$ matrix
\begin{displaymath}
  \underline{\underline{\chi}}_{\alpha,\beta}({\bf q})=
      \left(
      \begin{array}{cc}
        \chi^{\rho,\rho}_{\alpha,\beta}({\bf q}) & \chi^{\rho,m}_{\alpha,\beta}({\bf q}) \\
        \chi^{m,\rho}_{\alpha,\beta}({\bf q}) & \chi^{m,m}_{\alpha,\beta}({\bf q}) \\
      \end{array}
      \right)\,.
      \end{displaymath}
  
The total susceptibility matrix is then of dimension $2\lambda_{mag}\times 2\lambda_{mag}$ where $\lambda_{mag}$ is the magnetic periodicity (in units of the lattice spacing). The corresponding RPA equation reads
\begin{equation}\label{eq:rpa}
  \underline{\underline{\chi}}^{ee}({\bf q})
  = \underline{\underline{\chi^0}}({\bf q})
  + \underline{\underline{\chi^0}}({\bf q})\,
  \underline{\underline{V^{ee}}}({\bf q})\,
\underline{\underline{\chi}}({\bf q}) 
\end{equation}
with the interaction given by
\begin{displaymath}
  \underline{\underline{V}}_{\alpha,\beta}({\bf q})=
       \left(
      \begin{array}{cc}
        \frac{U}{2} & 0 \\
        0 & -\frac{U}{2}
      \end{array}
      \right)\delta_{\alpha,\beta}  \,.
\end{displaymath}

Upon including also the coupling to lattice fluctuations (vertex $g_{\bf q}$, frequency $\Omega_{\bf q}$)
the renormalized phonon propagator can be obtained from
\begin{displaymath}
  \underline{\underline{D}}({\bf q},\omega)=\left\lbrack
  \underline{\underline{1}}+ \underline{\underline{\Lambda}}({\bf q}) \underline{\underline{D^0}}({\bf q},\omega) \underline{\underline{\chi_{ee}}}(q,\omega)
  \underline{\underline{\Lambda}}({\bf q}) \right\rbrack^{-1} \underline{\underline{D^0}}({\bf q},\omega)
\end{displaymath}
where $\underline{\underline{\chi_{ee}}}({\bf q},\omega)$ is evaluated from Eq. (\ref{eq:rpa}). The vertex and phonon Greens function matrices are given by
\begin{eqnarray*}
  \underline{\underline{\Lambda}}_{\alpha,\beta}({\bf q})&=&
  \left(
      \begin{array}{cc}
        g_{{\bf q}+{\bf Q}_\alpha} & 0 \\
        0 & 0 
      \end{array}
      \right)\delta_{\alpha,\beta} \\
\underline{\underline{D}}^0_{\alpha,\beta}({\bf q})&=& \left(
      \begin{array}{cc}
        D_0({\bf q}+{\bf Q}_\alpha,\omega)& 0 \\
        0 & 0 
      \end{array}
      \right)\delta_{\alpha,\beta} \,
\end{eqnarray*}
with the bare phonon Green function
\begin{displaymath}
  D_0({\bf q},\omega)=\frac{2\Omega_{\bf q}}{\omega^2-\Omega_{\bf q}^2}\,.
\end{displaymath}

The phonon propagator can then be used to compute the phonon contribution to RIXS following the approach of Ref. \onlinecite{Devereaux2016}

In the main part of the paper we show results for a longitudinal acoustic phonon with frequency and coupling given by
\begin{eqnarray*}
  \Omega_{\bf q}&=&\Omega_0 \sqrt{\sin^2\frac{{\bf q}_x}{2}+\sin^2\frac{{\bf q}_y}{2}}\\
  g_{\bf q}&=&g_0 \sqrt{\sin^2\frac{{\bf q}_x}{2}+\sin^2\frac{{\bf q}_y}{2}}\,.
\end{eqnarray*}

The stripe phason mixes phonons which differ by a reciprocal lattice vector of the stripe lattice. This coupling is particularly strong at the stripe momentum ${\bf Q}_{\text{CDW}}=(0.25,0)$ where it can induce a quasicritical mode due to a change of the respective stability of bond- and site centered stripes similar to Ref.~\cite{Lorenzana2003a}. This mode is shown in Fig.~\ref{Fig1} (g,f)  for $\Omega_0/t=0.42$ and $g_0/t=0.73$.

We shall note that (i) $\Omega_{0}$ is the ``bare" phonon energy, which will be renormalized by the electron-phonon coupling  to an energy of order 70-80~meV for $t \sim 0.3$~eV; (ii) the optical phonon is also active in the energy range of interest, in this case the prominent asymmetric intensity distribution (Fig.~\ref{Fig2}a and b) is possibly caused by the more complicated cross-section effect of the RIXS process \cite{Ament2011}, which we did not take into account in our model calculations.

\section{Commensurate vs incommensurate CDW}

A commensurate CDW with period $Ma_{0}$ is know to have a strong lattice effect. Charge modulations mix electronic states with momentum $k+nQ_{\text{CDW}}$ ($n$ is an integer) and yields an additional phase dependent condensation energy \cite{Lee1974}. As temperature changes, this additional commensurate energy may thus drive an incommensurate to commensurate crossover. In mean field theory for a 1D CDW, the approximate crossover condition for $M=4$ is formulated as
\begin{equation}
|4a_{0}-\frac{1}{Q_{\text{CDW}}}|\leq\frac{2\pi^{2}}{\lambda_{\text{epc}}^{1/2}}\frac{E_{\text{cond}}}{D}
\end{equation}
where $E_{\text{cond}}=\frac{1}{2}n(\epsilon_{\text{F}})\Delta_{\text{CDW}}^{2}$ is the phase independent CDW condensation energy, $\Delta_{\text{CDW}}$ is the CDW gap, $D$ is the cut-off energy close to the bandwidth or Fermi energy and $\lambda_{\text{epc}}$ is a dimensionless electron-phonon coupling constant \cite{Gruner2018}. Evidence for this effect has been observed in conventional CDW materials, such as K$_{0.3}$MoO$_{4}$, TaS$_{3}$ and NdSe$_{3}$ \cite{Gruner2018}, where the CDW wavevector is temperature dependent and becomes commensurate at base temperature. This CDW evolution differs from our observations, where the commensurate CDW forms at high-temperature and persists to low-temperature.

It is worth to note that the pCDW is short-ranged without long range phase coherence. Since the multiple CDW periods are nearly degenerate in energy \cite{Corboz2014, Huang2017, Zheng2017}, it is reasonable to expect that multiple CDW periods coexist with $\lambda=4a_{0}$ being statistically dominated. This might be the reason of why $\mathbf{Q}_{\text{pCDW}}$ is slightly off 0.25~r.l.u. If possible, it would be interesting to directly check this in a future STM study.

\section{Entropy model for cuprate stripes}
\label{entropy}
Ishizaka et al \cite{Ishizaka2004} considered a successful model to explain the shift of incommensurability with temperature in nickelates. Here we first briefly review their model. They consider the entropy of doped holes as $S = k_{B}\text{ln}(N_{c}$) where $N_{c}$ is the number of configurations for a given concentration of holes. The number of configurations is computed as the number of ways to accommodate indistinguishable particles in boxes representing equivalent sites along the core of the domain wall that can accommodate holes. Nickelates have insulating stripes at $T = 0$. For filled stripes, there is only one configuration ($N_{c} = 1$) and $S = 0$. If the distance between the stripes is decreased at fixed $x$ there are not enough holes to fill completely all stripe core sites. Calling $\delta$ the concentration of electrons, the incommensurability, $\epsilon$, is now determined by the total number of “domain wall sites” or boxes being occupied by holes (concentration $x$) or electrons (concentration $\delta$), namely $\epsilon=x+\delta$. If there are $N$ total Ni sites in the system the entropy is
\begin{equation}
\begin{aligned}
S &=k_{B}\text{ln}\frac{N_{box}!}{N_{el}!(N_{box}-N_{el})!}\\&
=k_{B}N[\epsilon\text{ln}\epsilon-\delta\text{ln}\delta-(\epsilon-\delta)\text{ln}(\epsilon-\delta)].
\label{CDWentropy}
\end{aligned}
\end{equation}
Here we used the relation $\text{ln}N!=N\text{ln}N-N$ for $N\rightarrow\infty$. The computation is completed by postulating that the total energy is $E = \mu^{-} \delta N$ where $\mu^{-}$ is the energy to remove holes. Notice that this expression holds only for $\delta>0$. For hole addition a different energy is involved because the stripes are filled and the AF regions have to accommodate the holes. We call that energy $\mu^{+}$. The fact that $\mu^{-}$ and $\mu^{+}$ are different means simply that the filled stripe is an insulator and there is a jump in the chemical potential around $\delta=0$. To generalize this model to cuprates one should first identify the particles and the boxes. This is less trivial than in nickelates. For vertical stripes as in cuprates, if $d$ is the distance among domains (in units of the lattice constant $a$) the charge incommensurability is $\epsilon = 1/d$. The first model of stripes \cite{Zaanen1989} assumed insulating stripes as in nickelates and this leads to $\epsilon=x$. However, in the cuprates, $\epsilon = 2x$ is observed for $x<1/8$ \cite{Tranquada1995, Hucker2011} which leads to half-filled stripes. From the theory side, a more accurate computation \cite{Lorenzana2002} indeed predicted half-filled stripes in accord with experiment. On the other hand, metallic half-filled stripes pose a problem for the entropy model since the half-filled system has maximum entropy. Therefore, the temperature will only stabilize more this configuration and the incommensurability would be independent of temperature in contradiction with experiment. However, as we show in Fig.~\ref{Fig1}(a) the CDW is enhanced around $x = 1/8$. It was proposed by White and Scalapino \cite{White2003} that this 1/8-anomaly is due to the tendency of stripes to develop addition hole ordering along the stripe \cite{Lorenzana2002}. Indeed, assuming stripes in neighboring planes are perpendicular to each other, the Coulomb potential of one plane favors a half-filled CDW along the stripe in the next plane only at $x = 1/8$  consistent with the increased stability of the CDW at that doping. 

\begin{figure}[tb]
\includegraphics[width=8 cm]{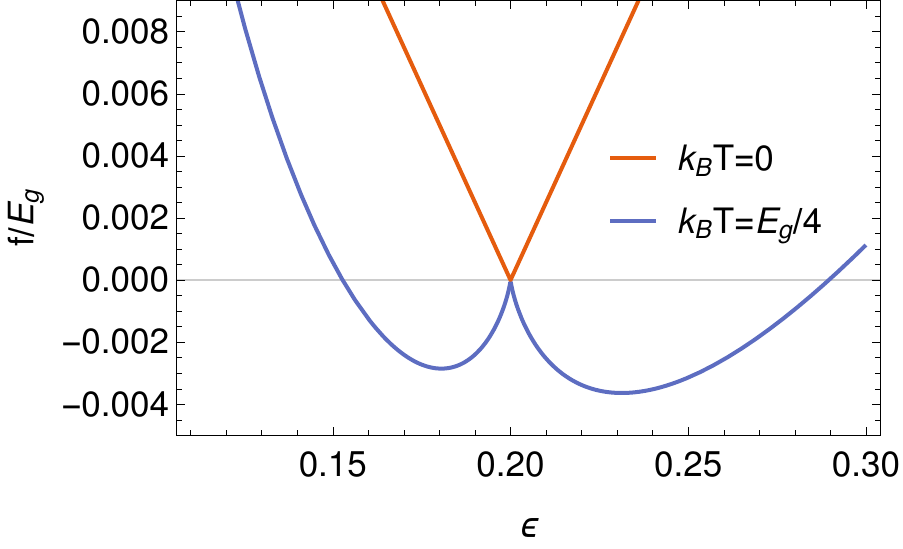}
\caption{\textbf{Free energy per Cu as a function of incommensurability}. Here we show the free energy of the entropy model at $x=0.1$ for two different temperatures (in units of the gap due to hole ordering along the charge stripe). At zero temperature the solution with $\epsilon=2x$ is favored while at finite temperature two values of $\epsilon$ minimize the energy with the one having  $\epsilon>2x$ prevailing.}
\label{Figure5}
\end{figure}

We can assume that decreasing the doping this configuration is still favorable as suggested by mean-field computations which picture the secondary CDW along the stripe as a lattice of Copper pair singlets\cite{Bosch2001}. In strong coupling, the pattern along the stripe is 00$\updownarrow\updownarrow$00$\updownarrow\updownarrow$, where 00 and $\updownarrow\updownarrow$ represent holes and disordered spins respectively. Notice that this has different periodicity than the pattern often assumed, $\updownarrow$0$\updownarrow$0$\updownarrow$.  Since at $T=0$ this state would be nominally insulating we assume again that the energy to add or remove hole is different, i.e.\ the pattern of site diagonal energy is assumed to be $\mu^{-}\mu^{-}\mu^{+}\mu^{+}\mu^{-}\mu^{-}\mu^{+}\mu^{+}$. As for the nickelates, this configuration has zero entropy. We now compute the entropy associated with an increase of the incommensurability i.e.\ a decrease of $d$ at fixed $x$. The incommensurability in this case satisfies $x + \delta = \epsilon/2$ where $\delta$ is the concentration of extra electrons. The entropy reads:
\begin{equation}
\begin{aligned}
S = k_{B}N\left[\frac{\epsilon}{2}\text{ln}\frac{\epsilon}{2}-|\delta|\text{ln}|\delta|-\left(\frac{\epsilon}{2}-|\delta|\right)\text{ln}\left(\frac{\epsilon}{2}-|\delta|\right)\right].
\label{CDWentropyb}
\end{aligned}
\end{equation}
where $N$ is the number of Cu sites and for simplicity we neglected the entropy due to the spin degrees of freedom which does not change the physics. 

If we consider a decrease of the incommensurability $x+\delta=\epsilon/2$ is still valid if we allow $\delta$ to be negative and interpret $-\delta$ as the concentration of holes added to $\mu^+$ sites. For fixed $\epsilon$, the entropy of the ordered half-filled stripe is symmetric with respect to adding or removing holes so Eq.~(\ref{CDWentropyb}) holds as written with the modulus and the energy can be written as $e = |\delta|E_{g}/2$ where we took $\mu^{\pm} = \pm E_{g}/2$ so the free energy is also symmetric. Notice however that the latter has to be minimized with respect to $\epsilon$ at fixed  $x$ which is not anymore symmetric. Indeed, there are two solutions which minimize the free energy at finite temperature as shown in Fig.~\ref{Figure5} having either $\epsilon<2 x$ or  $\epsilon>2 x$ and deviating from the zero temperature solution  $\epsilon=2 x$. The solution in which the incommensurability increases with temperature has lower free energy and leads to Eq.~\ref{CDWentropy1}. For $x>1/8$ the interaction between domain walls has to be taken into account\cite{Lorenzana2002}. A detailed theoretical study is left for future work as it goes beyond our present scope. In particular it would require adding additional terms to the energy that makes the low-temperature incommensurability to saturate at $\epsilon\sim 1/4$ and include the effect of Fermi surface wrapping which frustrates the secondary order along the stripe \cite{Anisimov2004}. For simplicity, here we neglect these effects and simply assume the solution in which the incommensurability decrease with temperature is favored due to the lower energetic cost. This leads to Eq.~\ref{CDWentropy2} for doping larger than 1/8. 

\begin{figure}[tb]
\includegraphics[width=8 cm]{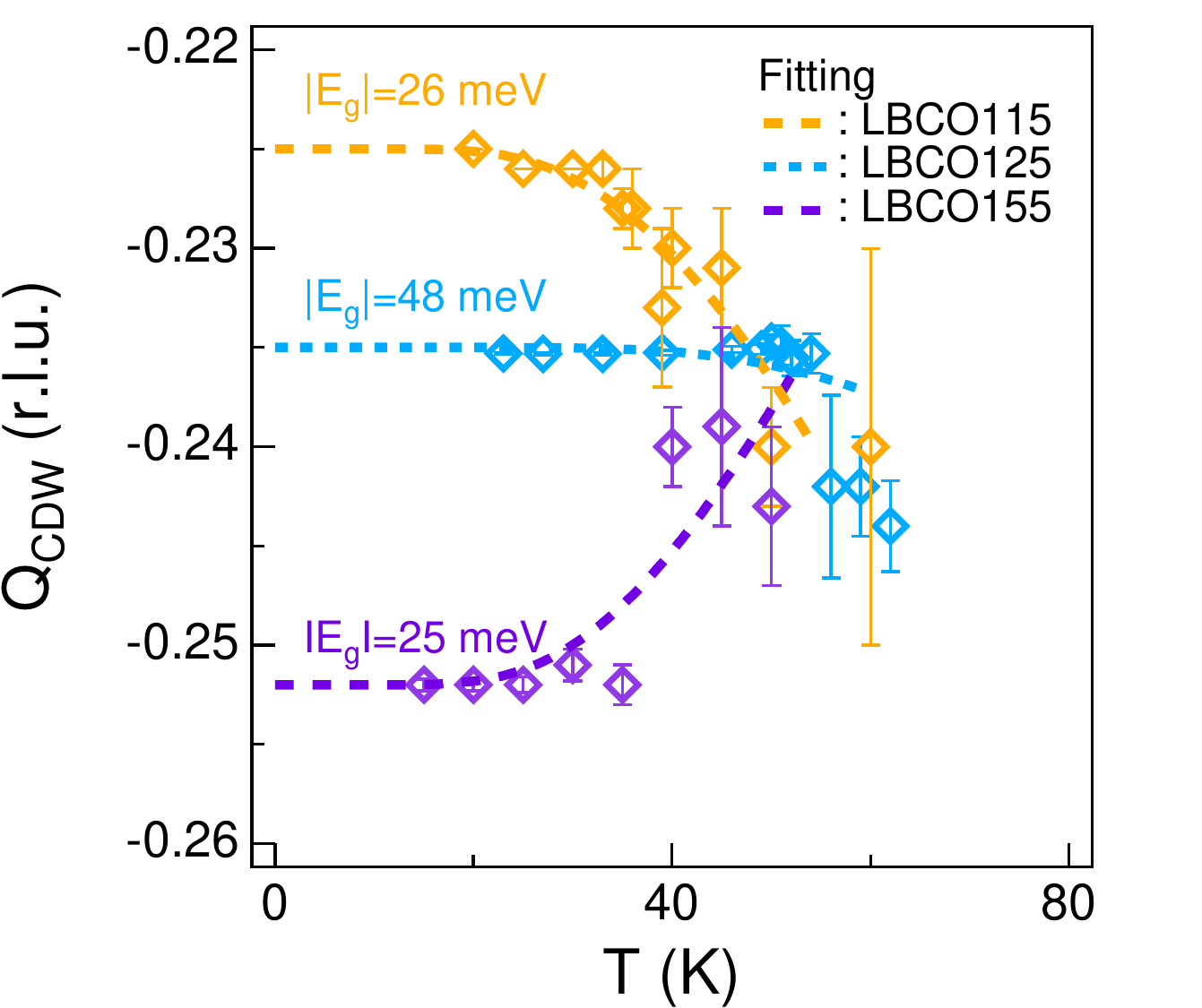}
\caption{\textbf{Entropy driven self-doping model}. Data in Fig.~\ref{Fig4}b are fitted by using Eq.~\ref{CDWentropy1} and Eq.~\ref{CDWentropy2}.}
\label{Figure6}
\end{figure}

We shall note our entropy model can qualitatively explain the incommensurate-commensurate crossover below or near the lCDW transition temperature, it, however, does not yield the saturation of $\epsilon_0$ as a function of doping or temperature due to the crude approximations. Figure~\ref{Figure6} shows a fit of experimental data below 60~K by using Eq.~\ref{CDWentropy1} and Eq.~\ref{CDWentropy2}. We also note that experimental studies of YBCO suggest that the local commensurate period is more consistent with $3a_{0}$ \cite{Wu2015}, which might be due to the special chain structure that favors a different period. A more sophisticated model that incorporating the unidirectional field may be needed to explain the result in YBCO.

\bibliography{ref}

\end{document}


\title{Supplementary Materials for: \\Formation of Incommensurate Charge Density Waves in Cuprates} 

\date{\today}



\date{\today}

\maketitle
\section{Curve fitting}

Here we present the fitting of the RIXS momentum scans integrated in an energy window of $\pm 100$~meV around $E_{\text{loss}}=0$~meV. The spectra are fitted by two Lorentzian squared functions plus a linear background:
%
%
\begin{equation}
\begin{aligned}
 I = c_{0}+c_{1}*x+P_{\text{I}}^{1}\left[\frac{1}{1+\left(\frac{x-P_{\text{pos}}^{1}}{P_{\text{wid}}^{1}}\right)^{2}}\right]^{2}+\\
 P_{\text{I}}^{2}\left[\frac{1}{1+\left(\frac{x-P_{\text{pos}}^{2}}{P_{\text{wid}}^{2}}\right)^{2}}\right]^{2},    
\end{aligned}
\label{Eq2}    
\end{equation}
where $P_{\text{I}}$, $P_{\text{pos}}$ and $P_{\text{wid}}$ represent peak intensity, position and width, respectively. Superscripts $1$ and $2$ differentiate the lCDW and pCDW peak. The linear term $c_{0}+c_{1}*x$ is used to account for the background. We chose the Lorentzian squared function on a phenomenological basis as it reproduces the observed peak shape better than other functions such as Lorentizian or Gaussian lineshapes \cite{Miao2017}. 

\begin{figure}[b]
\includegraphics[width=8 cm]{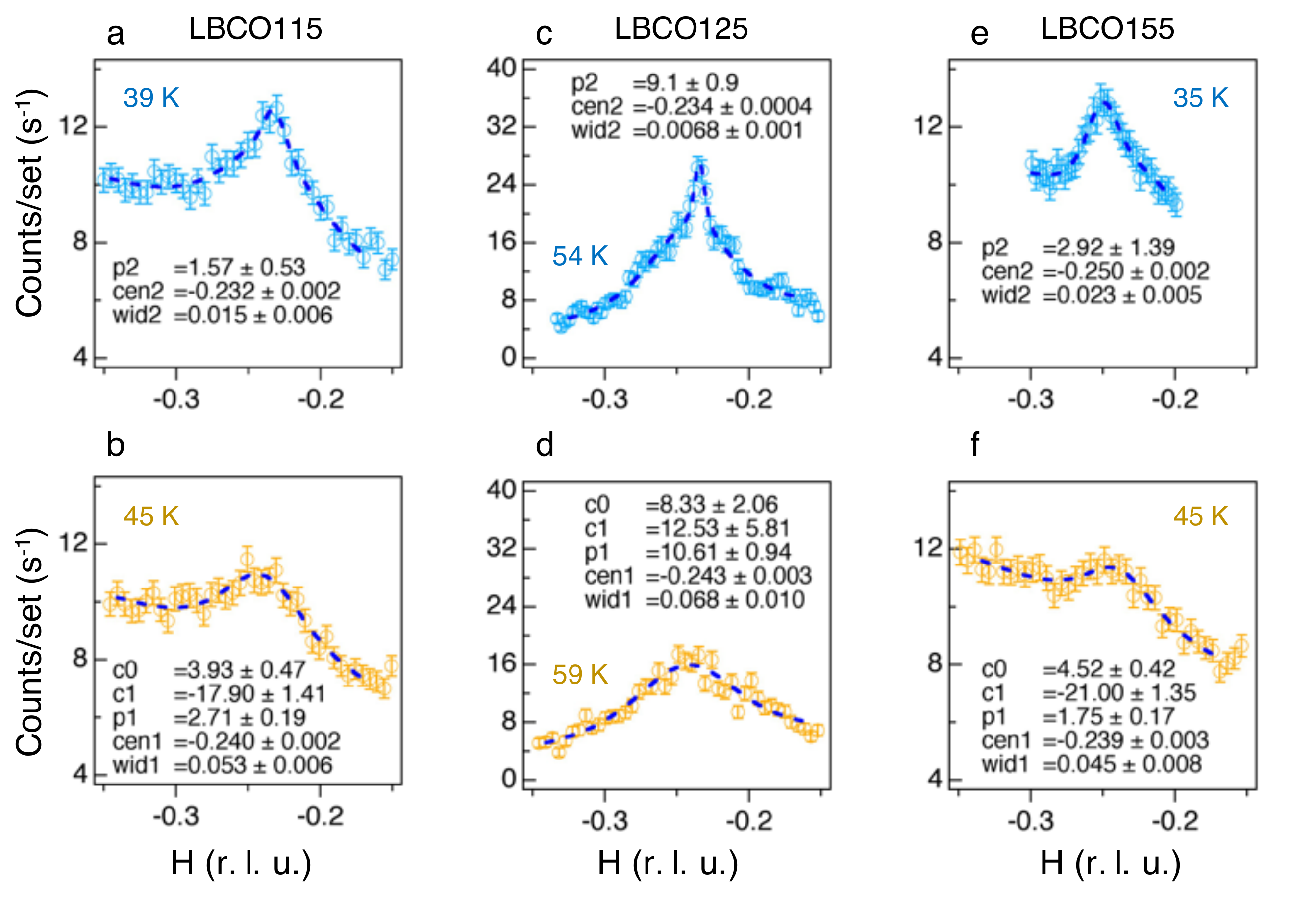}
\caption{\textbf{Curve fitting}. (a) and (b) show the fitted LBCO115 RIXS data slightly below (cyan) and above (yellow) the lCDW critical temperature, respectively. The same plots for LBCO125 and LBCO155 are shown in (c),(d) and (e),(f) respectively.}
\label{fitting}
\end{figure}

We start by fitting the pCDW peak at high temperature and set $P_{I}^{2}$ to be zero. Below the lCDW transition temperature, the integrated intensity has three contributions: lCDW, pCDW and a linear background. As shown in Fig.~3 of the main text, within experimental uncertainties, the shape of the broad intensity (that including both the pCDW peak and linear background) is nearly constant. We thus fix the fitting parameters of the pCDW and linear background in Eq.~\ref{Eq2}. Here the fixed fitting parameters are from data that just above the lCDW transition temperature. Figure~\ref{fitting} shows the fitted RIXS intensity near the lCDW critical temperature along the $H$ direction.

The volume of CDW is estimated by the integrated intensity defined as:
\begin{equation}\label{eq:OP}
  I_{\text{int}}=P_{\text{I}}*\text{HWHM}^{2}
\end{equation}
where $\text{HWHM}=\sqrt{\sqrt{2}-1}*P_{\text{wid}}$ is the half-width-at-half-maximum.
%

\bibliography{ref}